\documentclass[]{aa}  

\usepackage{graphicx}
\usepackage{txfonts}
\usepackage{hyperref}
\bibpunct{(}{)}{;}{a}{}{,} 
\newcommand{\itg}{{\it INTEGRAL~}}

\newcommand{\al}{{$^{26}$Al~}}
\newcommand{\HI}{\mbox{H{\hspace{1pt}}I}~}
\newcommand{\HIi}{\mbox{H{\hspace{1pt}}I}}
\newcommand{\HII}{\mbox{H{\hspace{1pt}}II}~}
\newcommand{\ali}{{$^{26}$Al}}
\newcommand{\msun}{{M$_\odot$}}
\newcommand{\mspy}{{M$_\odot$~yr$^{-1}$}}
\newcommand{\pccm}{{cm$^{-3}$}}
\newcommand{\kms}{{km~s$^{-1}$}}
\newcommand{\nbr}[1]{\left( #1 \right)} 

\newcommand{\rv}[1]{{#1}}
\newcommand{\fwidthfac}{0.48}
\newcommand{\capsize}{\tiny}

\begin{document} 

\title{\al kinematics: superbubbles following the spiral arms?}
\titlerunning{}
\subtitle{Constraints from the statistics of star clusters and \HI supershells}
 \author{Martin G. H. Krause \inst{1,2,6} \fnmsep\thanks{E-mail:
    krause@mpe.mpg.de} 
	\and Roland Diehl \inst{2,1} 
	\and Yiannis Bagetakos \inst{3,4}
	\and Elias Brinks \inst{5}
   	\and Andreas Burkert \inst{6,1,2}
	\and \\ Ortwin Gerhard \inst{2} 
	\and Jochen Greiner \inst{2}
	\and Karsten Kretschmer \inst{7}
	\and Thomas Siegert \inst{2}
}

\institute{
Excellence Cluster Universe, Technische Universit\"at
  M\"unchen, Boltzmannstrasse 2, 85748 Garching, Germany 
\and  Max-Planck-Institut f\"ur extraterrestrische Physik,
  Giessenbachstr.~1, 85741 Garching, Germany 
\and Kapteyn Astronomical Institute, University of Groningen, Postbus 800, 9700 AV Groningen, 
	The Netherlands
\and Netherlands Institute for Radio Astronomy, Postbus 2, 7990 AA, Dwingeloo, The Netherlands	
\and Centre for Astrophysics Research, University of Hertfordshire, Hatfield AL10 9AB, UK
\and Universit\"ats-Sternwarte M\"unchen, Ludwig-Maximilians-Universit\"at,
	Scheinerstr. 1, 81679 M\"unchen, Germany 
\and Fran\c{c}ois Arago Centre, APC, Universit\'e Paris Diderot,
 CNRS/IN2P3, CEA/Irfu, Observatoire de Paris, Sorbonne Paris Cit\'e,
 10 rue Alice Domon et L\'eonie Duquet, 75205 Paris Cedex 13, France
}

   \date{Received February 9, 2015; accepted ?}

 
  \abstract
   {\rv{High energy resolution} spectroscopy of the 1.8~MeV radioactive 
	decay line of \al with the 
	SPI instrument
	on board the \itg satellite has recently revealed that diffuse \al has large velocities in comparison to other components of the
	interstellar medium in the Milky Way. \al shows Galactic rotation in the same sense as the stars and other gas tracers, but reaches \rv{excess velocities} up to 300~km~s$^{-1}$.	}
   {We investigate if this result can be understood in the context of superbubbles, taking into account the statistics of young star clusters and \HI supershells, as well as the association of young star clusters with spiral arms.}
   {We derive energy output and \al mass of star clusters as a function of the cluster mass via population synthesis from stellar evolutionary tracks of massive stars. Using the limiting cases of weakly-dissipative and strongly-dissipative superbubble expansion, we link this to the size distribution of \HI supershells and assess the properties of likely \ali-carrying superbubbles.  
}
   {\al is produced by star clusters of all masses above $\approx 200$~\msun, 
roughly equally 
contributed over a logarithmic star cluster mass scale, and strongly linked to the injection of feedback energy. The observed superbubble size distribution cannot be related to 
the star cluster mass function in a straight forward manner. In order to avoid that the added volume of all superbubbles exceeds the volume of the Milky Way, individual superbubbles have to merge frequently.
\rv{If any two superbubbles merge, or if \al is injected off-centre in a bigger HI supershell we expect
the hot \ali-carrying gas to obtain velocities of the order of the typical sound speed 
in superbubbles, $\approx 300$~\kms before decay.} For star formation coordinated 
by \rv{the} spiral arm \rv{pattern} which, inside corotation, \rv{is overtaken by the faster moving} stars and gas, outflows from spiral arm star clusters \rv{would}
flow preferentially
into the cavities inflated by previous star formation associated with this arm. \rv{Such cavities would
preferentially be located} towards the leading edge \rv{of a given arm}.}
   {This scenario might explain the \al kinematics.
The massive-star ejecta \rv{are expected to survive} $\ge 10^6$~yr before being recycled 
into next-generation stars.}

   \maketitle
%
\defcitealias{Kretschea13}{Paper~I}

\section{Introduction}\label{sec:intro}

The possibilities to obtain kinematic information for the hot phase of the 
interstellar medium (ISM) are generally very limited. While multimillion-degree
gas is common, and metal lines are observed \citep[e.g.,][]{HS12}, the spectral 
resolution typically does not allow \rv{one} to meaningfully constrain flows of hot gas in galaxy clusters
\citep{BDB13}, and more so for the smaller velocities in the ISM.

The gamma-ray spectrometer aboard \itg \citep[SPI,][]{Vedrea03,Winklea03}, 
has a spectral resolution of $\approx$~3~keV at 
1.8~MeV, where \al can be observed through its characteristic
gamma-ray decay line. With increasing exposure times
\citep[Paper~I in the following]{Diehlea06,Kretschea13}, it has become 
possible to measure the centroid 
position of the line with an accuracy of tens of \kms, sufficient to clearly observe
the Doppler shift due to large-scale rotation along the ridge of the Galaxy
within longitudes $|l|<35$~deg. 

Towards the Galactic 
centre ($l=0$), the apparent \al velocity is zero with a hint for a small blue-shift. For
greater positive (negative) longitudes, the projected velocity 
rises beyond 200 (-200)~\kms. The direction of the line shift corresponds 
to 
Galactic rotation, but its magnitude is significantly larger
than what is expected from CO and \HI. \citetalias{Kretschea13} also showed that an ad hoc model
assuming forward blowout at 200~\kms~from the spiral arms of the inner Galaxy can well explain the data. 
The physical interpretation 
would be that \al is ejected into \rv{the hot phase of the ISM in} 
superbubbles at the leading edges of the gaseous
spiral arms. Hydrodynamic interaction with the locally anisotropic ISM
would then lead to a preferential expansion of the superbubbles into the 
direction of Galactic rotation  (in addition to out-of-plane-blowout).

The sources of \rv{diffuse, interstellar} \al are massive star winds and supernovae \citep{PD96}.
These are energetic events, which lead to the formation of bubbles 
\rv{(one massive star)}
and, because massive stars \rv{often} occur 
\rv{together with other massive stars in associations and bound clusters, }
\citep[e.g.,][]{ZY07,Kroupea13,Krumholz14}, 
superbubbles. Superbubbles are observed in many different wavelengths
\citep[e.g.,][]{Krausea14a}. Statistical information is, however, mainly restricted
to sizes and kinematics of the cavities seen in \HIi.
\rv{\citet{Bagea11} analysed 20~nearby spiral galaxies, whose properties are thought to be similar to those of the Milky Way, and found more than 1000 "\HI holes". We use their data
as reference below.} 
\citet{OC97} have connected the statistics of \HI holes to the star cluster mass function,
finding that the sizes and velocities of \HI holes may be explained 
by massive star activity in star clusters (compare below, however). 
\rv{Because this association is established now, we will in the following 
use the term "HI supershells" instead of "HI holes", for clarity.}

The \al measurement constitutes another piece of statistical information for 
\rv{bubbles and} superbubbles. 
\al decays on a timescale of 1~Myr\rv{, much less than typical superbubble lifetimes
\citep[e.g.,][]{OG04,Bagea11,Heesea15}}. Hence, we may expect it to reflect internal dynamics. 

Here, we connect the observed \al
kinematics to the statistics of star clusters (Sect.~\ref{sec:sc-al}) and superbubbles 
(HI supershells, Sect.~\ref{sec:sc-sb})\rv{, in order to better understand the large-scale 
gas flows traced by \ali}. 
\rv{In particular, we are interested to constrain superbubble merging, because superbubble 
merging may lead to asymmetric motions relative to the parent star clusters, when gas
from a high pressure superbubble streams into a low-pressure cavity.}
We find that star clusters of all masses contribute 
to the \al signal. 
\rv{\citet{OC97} investigate superbubble merging in the Milky Way with 
inconclusive results.
With updated models and star-formation rate we find frequent merging.
Hence, we expect the \ali-traced
hot outflows} to be injected into pre-existing superbubbles. 
We then argue in Sect.~\ref{sec:concmod} that the spatial co-ordination of star formation in the Milky Way by
the spiral arms may lead to the observed \al kinematics.

\section{Which star clusters produce how much \ali?}
\label{sec:sc-al} 

\rv{Star formation generally takes place in clusters and associations,
the majority of which disperse after some time
\citep{LL03,Kruijssen12}. For the case of bound star clusters 
it is debated if the dispersal is 
due to gas expulsion \citep[e.g.,][]{GiBa08}. Recent observations did
not find the expansion velocities expected if gas expulsion was important
\citep[e.g.,][]{HenBruea12}. Hence,
the dispersal is probably related to tidal effects \citep[e.g.,][]{Kruijea12a},
which could take as long as 200~Myr \citep{Kruijea12b}. It is therefore
reasonable to assume that for the timescales of interest here, 
the great majority of massive stars are grouped \citep[compare also][]{ZY07}.
The mass function of embedded star clusters (mostly unbound), which is where
most star formation takes place locally \citep{LL03} has a very similar slope
than the one of star clusters in external galaxies (compare below). Therefore,
we assume just one mass function for star forming regions in the following,
and generally use the term "star cluster" without qualifying adjective
to subsume bound and unbound star forming regions.  }

For spiral galaxies like the Milky Way, the initial cluster mass function (ICMF) is given
by \citep[e.g.,][]{Larsen09,Bastea12}
\begin{equation}\label{eq:ICMF}
\frac{\mathrm{d}N}{\mathrm{d}M} = a \nbr{M/M_\mathrm{c}}^\alpha
	\exp\nbr{-M/M_\mathrm{c}} \, ,
\end{equation}
where $a$ is the normalisation, the cutoff mass 
$M_\mathrm{c}= 2 \times 10^{5}$~\msun, and we take the
power-law index $\alpha$ to be $-2$; 
compare also the reviews
by \citet{LL03,Kroupea13,Krumholz14}. Following \citet{LL03}, we
adopt a lower limit for star cluster masses of 50~\msun.
\rv{Embedded star clusters have not been shown to possess the exponential cutoff.}
We have\rv{, therefore,} checked that the \rv{presence} of the high-mass cutoff only marginally influences our results. 

Since only massive stars produce \ali, we have to relate the occurrence of massive 
stars to the masses of star clusters. 
We \rv{carry out the entire analysis for both, optimal sampling \citep{Kroupea13},
where the masses of massive stars are fixed for given star cluster mass,
and random sampling \citep[e.g.,][]{Krumholz14}. For random sampling, we 
fix the stellar mass above 6~\msun~to the corresponding fraction of the IMF from \citep{Kroupea13}.
While the extreme assumption of
optimal sampling has been challenged recently \citep{Andrewsea14}, we use it here
to demonstrate that even such a strong truncation of the IMFs would not affect
the conclusions.}


\rv{For such groups of massive stars, w}e use the population synthesis results from \citet{Vossea09} (stellar evolutionary tracks of rotating
stars of \citet{MM05} and wind velocities from \citet{Lamea95} and \citet{NiSk02} 
for the Wolf-Rayet phase) to obtain the \al mass as well as the energy
injected into the ISM by massive stars as a function of time and stellar 
mass. The release of mass, energy and \al
is largely completed after about 48~Myr, the lifetime of stars of about 8~\msun,
also broadly consistent with the age estimates
for \HI supershells given by \citet{Bagea11}.
Not all the stars in a cluster might form at the same time. However typical age spreads within clusters are of order 1~Myr or below \citep[e.g.,][]{Niedea15},
which is much shorter than the timescales of interest.
We therefore use the star cluster population up to 48~Myr for our model.
Following \citet{CP11}, we take 1.9~\mspy~
for the star formation rate of the Milky Way. 
This \rv{sets} the constant in eq.~(\ref{eq:ICMF}) to $a=3\times10^{-4}$~\msun$^{-1}$.
Uncertainties in this parameter are substantial
\citep[compare also][]{KE12}. 

\begin{figure}
  \centering
  \includegraphics[width=\fwidthfac\textwidth]{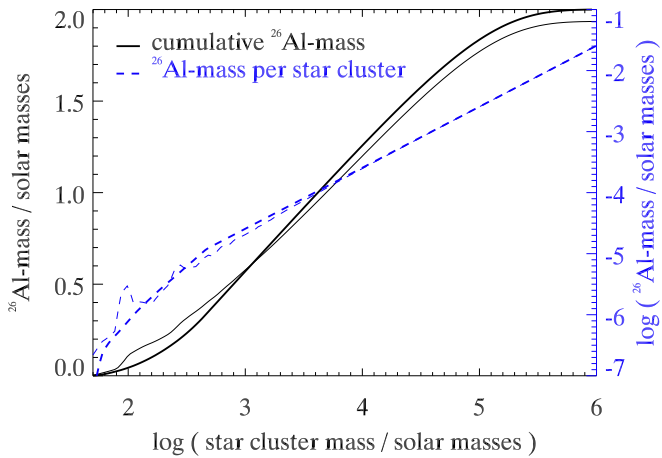}
  \caption{\capsize \al mass for individual star clusters of given mass (dotted blue, right vertical scale) and 
	cumulative \al mass for the Milky Way as a function of star cluster mass, 
	assuming a star formation rate of 1.9~\mspy (solid black, left vertical scale). 
	\rv{Thin (thick) lines are for the case of truncated IMFs (random sampling). 
	In the limit of high star cluster masses, $2.6 \times 10^{-8}$~\msun~of \al is 
	produced per unit stellar mass formed. The blue dashed curves are therefore 
	linear, down to about 1000~\msun, where sampling effects become important.}}
   \label{f:al_per_sc}%
\end{figure}

With these assumptions, we calculate the time-averaged \al mass for a given star cluster.
For each star cluster, we first determine the masses of its
stars above 8~\msun~by the optimal sampling method, the amount of released
\al from \citet{Vossea09}, taking into account radioactive decay, and finally average over time (48~Myr). 
The result is shown in Fig.~\ref{f:al_per_sc}. Apart from small
features towards lower masses, the \al yield is almost linear 
\rv{even for optimal sampling. For star clusters below about 1000~\msun,
the sampling method matters. For both, random and optimal sampling, the 
\al mass per cluster drops below the linear relation, because the IMF can no longer 
be fully sampled (e.g. a 120~\msun~star may not live in a 50~\msun~star cluster).}
\rv{We note that using the IMF directly, without dividing the mass of young stars into star clusters, to predict the Galactic \al mass yields a
higher value by about 20 per cent.}

The ICMF has roughly equal mass in each decade of star cluster mass 
(within the cutoffs). This remains true with the \al mass folded in, because 
the latter is roughly proportional to the star cluster mass: star clusters of 
each decade in mass\rv{, from a few hundred to about $10^5$~\msun,}
contribute about equally to the observed \al signal (Fig.~\ref{f:al_per_sc}). 


\section{Superbubble size distributions and merging}
\label{sec:sc-sb}
\begin{figure}
  \centering
  \includegraphics[width=\fwidthfac\textwidth]{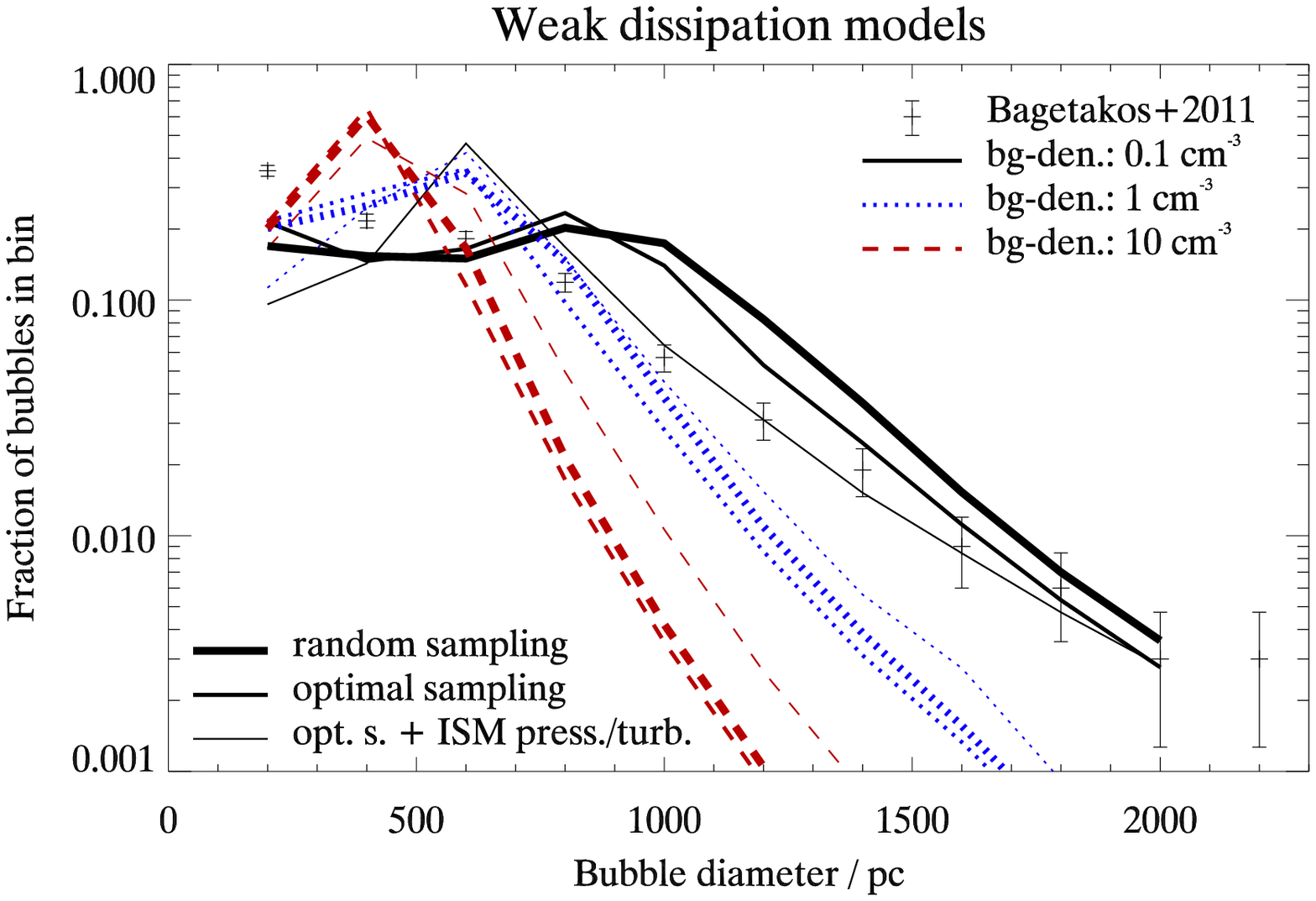}
  \includegraphics[width=\fwidthfac\textwidth]{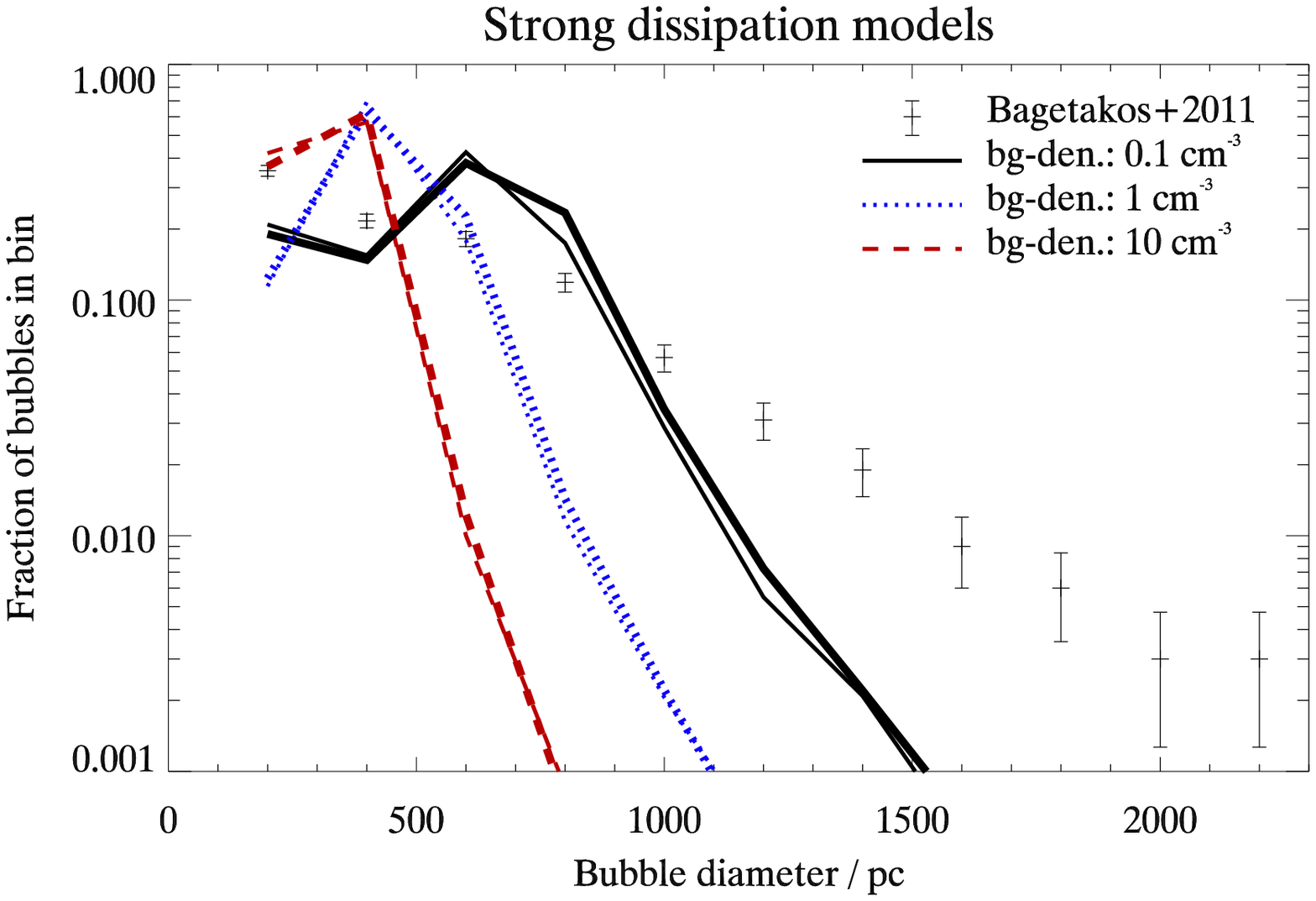}
  \caption{\capsize Superbubble diameter distributions for the weakly (top) and the strongly (bottom)
dissipative model for three different choices of the background density (bg-den. in the legends). The size of the bins is 200~pc. \rv{Thick lines are for random sampling,
thinner ones for optimal sampling, and the thinnest ones in the top panel 
are for optimal sampling where the background pressure and superbubble destruction by ISM turbulence are taken into account.}
The minimum near 400~pc for the solid curves is due to the strong acceleration after the first supernova in a superbubble. It is below the data range for the other curves. Large superbubbles are better explained by the weakly dissipative model.}
   \label{f:hi_hole_size}%
\end{figure}
\rv{Here, we investigate if merging of superbubbles is common in the Milky Way.}
\rv{We follow the overall procedure described in \citet{OC97}, but update 
the expansion models 
\rv{from our own 3D hydrodynamics simulation studies}
(compare below). While towards the low mass end, the ICMF includes many 
objects with only one massive star, which will produce a single-star bubble, 
we use the term
"superbubble", below for simplicity for all bubbles produced by the star clusters.}

\citet{McLMcC88}
present a self-similar model for superbubble expansion, where the superbubble expands
steadily with radius $r$ proportional to a power law in time $t$. About 35 per cent of 
the injected energy, $E(t)$, is dissipated radiatively in this model. This model should be increasingly
adequate for larger superbubbles, with more frequent explosions, and at later times. 

In \citet{Krausea13a} and 
\citet{KD14}, we have developed a more strongly dissipative model from 3D
hydrodynamics simulations. The reason for the stronger dissipation is the 
more realistic, non-steady energy input and the emergence of a highly radiative mixing layer due to 3D instabilities.
Our results are well approximated by 90 per cent dissipation in the
steady energy input phase before the first supernova and a decline of the current energy, $E(t)$, after each supernova with time $t$ as  $t^{-3/4}$
\rv{(momentum-conserving snowplough)}. Both are an upper limit on the 
energy dissipation, as in the pre-supernova phase we still observed a slight  
dependence on numerical resolution ($\approx 88$ per cent dissipation at the highest resolution) and, as the superbubble expands, the
density around star clusters will drop below the 10~\pccm~we assumed 
in the simulations. 
The strongly dissipative model should be more adequate for superbubbles
with few supernovae,
and indeed explains, e.g., the X-ray-luminosity--kinematics relation well \citep{KD14}.

We use the evolution of the superbubble energy $E(t)$ from both models and 
predict the radius in the thin shell approximation 
following \citet{KD14}. Their eq.~(3) for constant \rv{ambient} density
$\rho_0$ evaluates to
\rv{$r^5 = 15/(2\pi\rho_0) \int_0^t dt^\prime \int_0^{t^\prime} dt^{\prime\prime} E(t^{\prime\prime}).$}

\rv{We calculate models for both, random sampling and optimal sampling.
For the weakly dissipative models, we also add models where we take a constant 
ISM pressure of $P_0=3800 k_\mathrm{B} \mathrm{K} \mathrm{cm}^{-3}$
\citep{JT11} into account which limits the expansion. The momentum 
equation may then be written as \citep{Krause2005a}:
$\partial^2Y(r)/\partial t^2 = E(t) - 2 \pi r^3 P_0$, with 
$Y(r) = 2\pi \rho_0 r^5 / 15$, which we solve numerically.
For this model, we also
regard a superbubble as dissolved when the expansion velocity has dropped 
to 10~km~s$^{-1}$ and perturbations with this velocity had time to grow
to the size of the superbubble, similar to the "stalled and surviving" mode 
in \citep{OC97}. We do not investigate this option for the strong dissipation models,
because the assumption of momentum conservation after each supernova explosion 
implies a total pressure force of zero.}

For the following analysis, we neglect the shear gradient from galactic rotation. 
It is typically 10-50~\kms~kpc$^{-1}$ \citep{Bagea11}, 
and therefore has a small effect on active superbubbles, in agreement with the moderate 
asymmetries found by \citet{Bagea11}, but will eventually destroy old ones.
The finite exponential scaleheight $H$ of the ISM introduces a cutoff in the superbubble radii
in the Galactic plane
due to blow-out related pressure loss at $\approx 3H$ \citep{BB13}. We set this cutoff superbubble radius to  1~kpc for the whole sample, and 0.5~kpc for the Milky Way modelling below due to the lower \HI scaleheight \rv{\citep{NJ02,LPV14}}.

\rv{We first calculate the fractional distributions of superbubble diameters 
for three different 
assumptions for the background density
for the sample of star-forming galaxies from \citet{Bagea11}, i.e. $H=1/3$~kpc,
and compare this to the observations in Fig.~\ref{f:hi_hole_size}.} 
Generally, models with lower background density provide a better match to the observations.
As expected, the weakly dissipative model more closely represents the large superbubbles. 
The model is \rv{not quite} satisfactory, \rv{because}
the density \rv{required to reach the larger diameters}, 0.1~\pccm, is on the low side of values suggested by observations,
0.1-0.7~\pccm \citep{Bagea11}. One might be able to interpret this finding by 
shear effects, adopting a higher density, i.e. choosing
a curve between the dotted blue and solid black lines in Fig.~\ref{f:hi_hole_size}. 

\rv{The strongly dissipative models may produce
a significant population at around 1 kpc diameter, but, on the other hand,
cannot account for large \HI supershells.}
\rv{ISM pressure becomes most important for intermediate-size \HI supershells 
and for low ISM density
($\approx 1$~kpc for $\rho_0=0.1$~cm$^{-3}$). At high ISM densities, ISM pressure is negligible, but in these models
many slower and smaller superbubbles are destroyed when considering ISM turbulence,
which increases the fraction of larger superbubbles.}
\rv{The IMF sampling method has a minor effect on the results 
(compare Fig.~\ref{f:hi_hole_size}).}

\rv{We can now predict the superbubble distribution for the Milky Way from the star formation rate using the procedure outlined above, now with $H=1/6$~kpc.
The fractional distributions are identical to Fig.~\ref{f:hi_hole_size},  but cut 
at 1~kpc due to the reduced scaleheight. The observed fractional \HI supershell
diameter distribution for the Milky Way \citep{EP13} is consistent with the one 
of external star-forming galaxies from \citep{Bagea11}, which we used here.}

\rv{Because for the Milky Way, the total number of superbubbles is constrained by the star-formation rate, we 
may now check for superbubble merging by calculating the total volume predicted by our model to be occupied by superbubbles and comparing it to the volume of the Milky Way ISM.
The} total 
occupied volume for the given star formation rate exceeds the one of the Milky Way ISM 
(cylinder: 10~kpc radius, 1~kpc thickness) for all assumptions (Table~\ref{t:vols}).

\rv{This indicates that the superbubbles merge frequently. In the case of merging superbubbles,
the total volume is not simply the sum of the individual volumes, but much smaller.
The observed volume fractions of \HI supershells (3D porosity) are typically below 10 per cent
and may reach 20 per cent in later Hubble types \citep{Bagea11}. 
A superbubble volume fraction around 20 per cent is expected from the hot gas fraction 
in the ISM simulations of \citet{dAB05}. Combined
with our analysis, this strengthens the point about superbubble merging. A consistent
interpretation would be that the smaller superbubbles in the diameter distribution
(Fig.~\ref{f:hi_hole_size}) merge to obtain more HI supershells at large diameters. This would 
also alleviate the requirement for low ambient density (compare above).
}


\rv{There is a lot of direct evidence for superbubble merging in the Milky Way:
29 per cent of the bubbles identified in "The Milky Way Project", a citizen science project
that identified 5106 bubbles in the Milky Way (many of which are single star bubbles),
showed signs of merging \citep{Simpsea12}. Often, secondary bubbles are found on the edge of larger bubbles. \citet{EP13} calculate the porosity for the Milky Way as a function
of radius from 333 identified HI supershells. 
They find porosities above unity inside of the solar circle, and thus strong overlap
of superbubbles. The closest massive star group, \object{Scorpius-Centaurus OB2},
is an excellent example for superbubble merging \citep{Poepea10,PreibMam08}:
the different subgroups of the OB association appear to have been triggered by expanding
shells from the older parts, and the shell around Upper Scorpius is half merged into an older supershell. The whole structure is expected to merge within a few Myr with the Local Bubble \citep{BdA06}. Evidence for superbubble merging from extragalactic studies 
is, however, scarce, probably because of the low resolution (typically around 200~pc).
H$\alpha$ bubbles are however found at the rims of HI supershells \citep{Egorea14}. }

\rv{Superbubble merging may produce significant net velocities in ejecta flows 
with respect to the driving massive-star group. Because the \al content
is correlated with the energy content of a superbubble (Fig.\ref{f:al-en}),
we expect overpressured \ali-rich material to often stream into lower pressured 
superbubbles, once the interface is eroded. The situation is similar, if the \al
production site is located towards one end of an already merged larger superbubble.}

%
%
\begin{table}
  \caption{\rv{Galaxy integrated superbubble volumes in units of the Milky Way volume, assuming  a maximum superbubble diameter of 1~kpc due to blowout. For each entry, the first (second) number is for random sampling (truncated IMFs). For weak-dissipation
models, we also give the numbers for the models that take into account the ISM background pressure and turbulence as the third number.}}             
  \label{t:vols}      
  \centering          
  \begin{tabular}{l r r r }     
    \hline\hline       
    Dissipation & $\rho_0=0.1$~\pccm& $\rho_0=1$~\pccm& $\rho_0=10$~\pccm\\
    \hline    
    weak & 115/115/33 & 40/38/22 & 11/11/6.1 \\
	strong & 47/49 & 12/13  & 3.1/3.3 \\
   \hline                  
  \end{tabular}
\end{table}
%

\section{A model for the \al kinematics}
 \label{sec:concmod}
\begin{figure}
  \centering
  \includegraphics[width=\fwidthfac\textwidth]{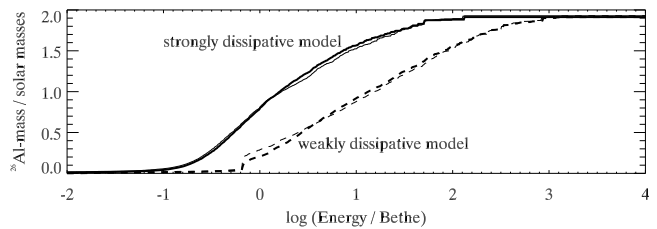}
  \caption{\capsize Cumulative \al mass over current superbubble energy for weakly (dotted) and strongly (solid)
dissipative models for a star cluster population representative of the Milky Way. 1 Bethe = $10^{51}$~erg.}
   \label{f:al-en}%
\end{figure}
\begin{figure}
  \centering
  \includegraphics[width=.4\textwidth]{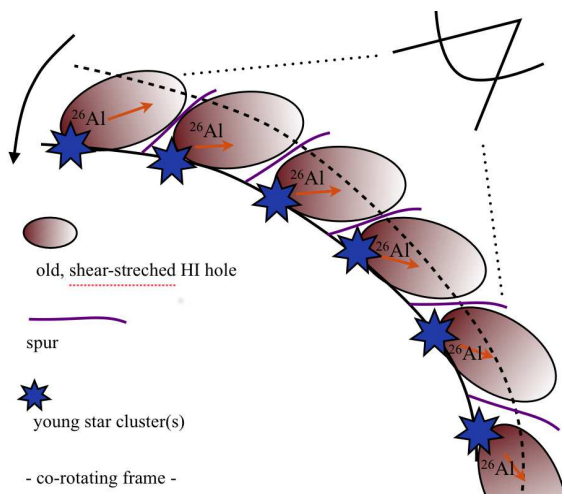}
  \caption{\capsize Sketch of the proposed model to explain the \al kinematics. In the co-rotating
frame chosen here, a spiral arm (solid line) moves anti-clockwise. At its 
previous location (dashed line), it created large superbubbles (ellipses), blowing out of the disc. The young
star clusters (blue stars) at the current spiral arm location feed \al (colour gradient in ellipses) 
into the old superbubbles.}
   \label{f:sketch}%
\end{figure}

In the preceding sections, we have demonstrated that star clusters of all masses 
are equally important as \al producers, and that, on Galactic scales, star clusters
cannot be assigned to individual superbubbles due to frequent superbubble merging.
Our model also shows that \al injection from star clusters is strongly correlated 
to energy injection (Fig.\ref{f:al-en}). It follows that \al is likely to be observed
in motion, and in particular it is likely that it traces gas involved in superbubble merging.
Based on these findings, we suggest the following model (Fig.\ref{f:sketch}) to explain 
the \al kinematics.
 
When spiral arms sweep through
the Galactic disc, they trigger the formation of young star clusters that produce 
large superbubbles, \rv{traced as} \HI supershells. During the observed lifetimes of \HI supershells,
$\lesssim 100$~Myr \citep{Bagea11}, a spiral arm may lag behind stars and gas by 
as much as a few kpc, due to the pattern speed
\rv{of the arm which is lower within corotation than the rotational speed of the stars and gas}.
The current young star clusters in a spiral arm
therefore feed \ali-carrying ejecta into the \HI supershells left behind by the receding spiral arm
(sketch in Fig.~\ref{f:sketch}). 
%

\rv{Despite uncertainties regarding wind clumping \citep[e.g.,][]{Bestenlea14} and 
dust production and clumping \citep[e.g.][]{Indea14,Williams14}, the bulk of \al is 
likely mixed into the diffuse gaseous ejecta, expelled into the hot immediate surroundings of the stars. The ejecta} do not keep their initial velocity ($\approx 1000$~\kms) for long: 
for supernovae, they are 
shocked on timescales of $10^3$~yr \citep{Tenea90}. For Wolf-Rayet winds inside 
superbubbles, the free expansion phase can be up to $10^4$~yr, or 
$\approx 10$~pc \citep{Krausea13a}.
The ejecta then travel at 
\rv{a reasonable fraction of}
 the sound speed in superbubbles,
$c_\mathrm{s} = \sqrt{1.62 kT/m_\mathrm{p}} = 279\,T_{0.5}^{1/2}$~\kms.  
Here, $k$ is Boltzmann's constant, $m_\mathrm{p}$ the proton mass, $T$ ($T_{0.5}$) 
the temperature (in units of 0.5~keV), and the numerical factor is calculated for a 
fully ionised plasma of 90~per cent hydrogen and 10~per cent helium by volume. 
Measurements of superbubble temperatures range from 0.1~keV to about 1~keV
\citep[e.g.,][]{DPC01,Jaskea11,Sasea11,KSP12,Warthea14}, in good agreement with expectations, if instabilities and mixing are taken into account \citep{Krausea14a}.

\rv{In simulations of merging bubbles \citep{Krausea13a}, we find such kinematics for gas flooding 
the cavities at lower pressure
shortly after merging. T}he ejecta travel about 300~pc during one 
decay time ($\tau = 1$~Myr), which 
corresponds to the size of the smaller \HI supershells (Fig.~\ref{f:hi_hole_size}),
i.e. the decay is expected to happen during the first crossing of the \HI supershell.

Hence, we expect a one-sided \al outflow at the superbubble sound speed, $\approx 300$~\kms,
in excellent agreement with the observations and their analysis presented in \citetalias{Kretschea13}.

\rv{This} model
predicts a change in \rv{relative} outflow direction near the corotation radius.
\rv{But, c}orotation \rv{in the Galaxy} is unfortunately too far out 
\citep[8.4-12 kpc, e.g.,][]{MBP15}
to check for direction reversals in the data set of \citetalias{Kretschea13}. 
At such galactocentric distances,
individual \ali-emission regions are \rv{only a few, faint, and} not associated with
spiral arms. Thus, we do not expect large \al velocity asymmetries, 
in good agreement with the measurements in Cygnus \citep{Martinea09}
and Scorpius-Centaurus \citep{Diehlea10}.

\begin{figure}
  \centering
  \includegraphics[width=\fwidthfac\textwidth]{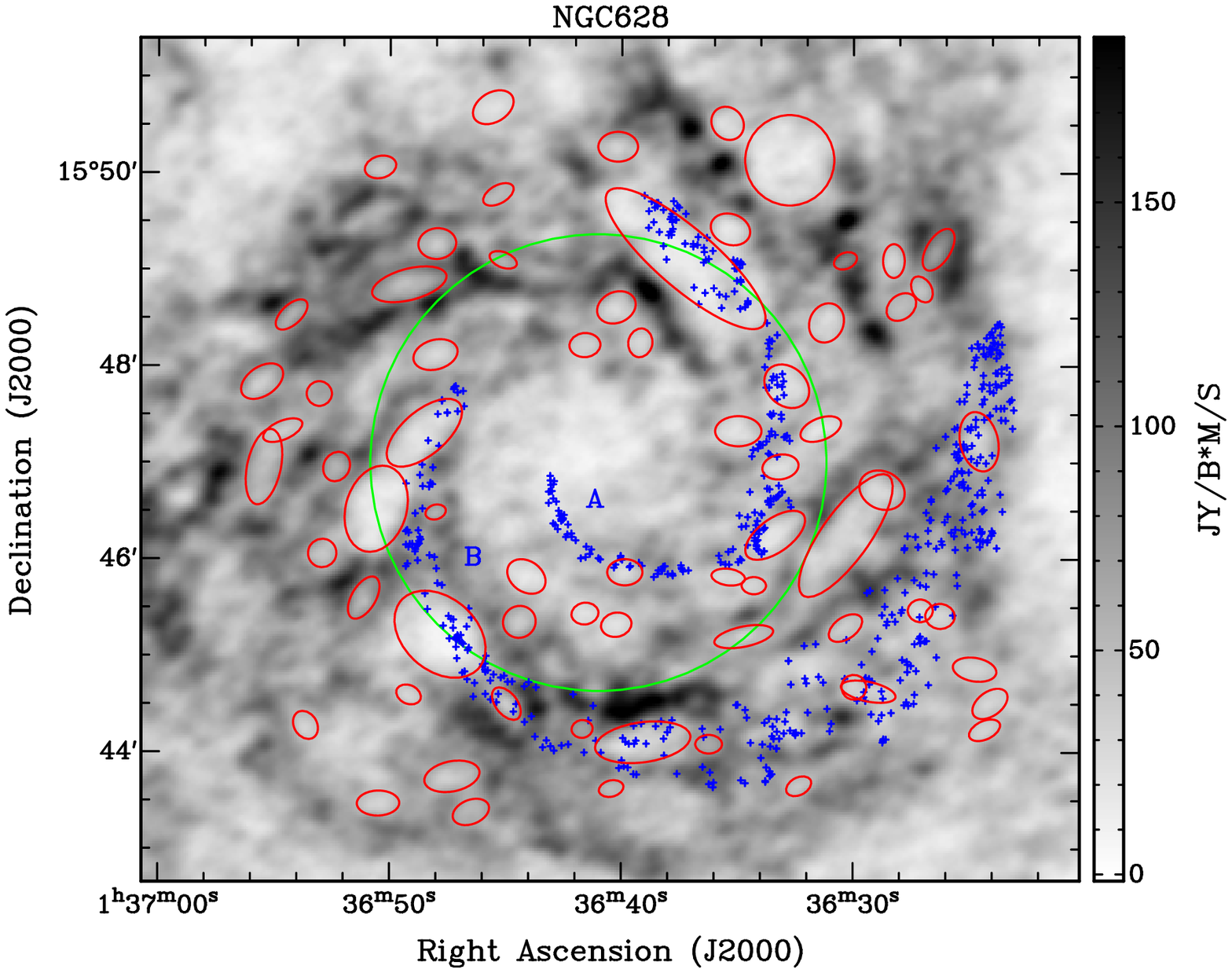}
  \caption{\capsize 
	The grand-design spiral galaxy \object{NGC~628}. The background 
	image is the 21~cm map from The \HI Nearby Galaxy Survey
	\citep[THINGS, ][]{Waltea08}. Red ellipses denote \HI supershells from
	\citet{Bagea11}. Blue 'plus'-signs denote the 650~\HII regions identified
	by \citet{HR15}. Their spiral arm designations, 'A' and 'B', are also indicated.
	The large green circle indicates the median corotation radius of 
	$4.6\pm1.2$~kpc
	from a number of studies as compiled by \citet{SL13}.
	For the first half-turn, arm 'A' has no \HI supershell on its trailing edge, but four
	are close to or even overlapping the leading edge in the way envisaged by our model.
	Arm 'B' begins just inside of corotation and has three prominent \HI supershells 
	at its leading edge, with only a minor one towards the trailing edge. 
	From about the corotation radius outwards, \HI supershells are no longer 
	at the edges of the \HII arm, but appear all over it.}
   \label{f:NGC628}%
\end{figure}

\rv{We might, however, expect to find \HI supershells associated with the
leading-edge of spiral-arm star-formation regions
in nearby face-on spiral galaxies, inside their corotation radii. We have 
investigated this for a few objects by combining \HII regions from 
\citet{HR15} to \HI images with \HI supershells using corotation radii from
\citet{Tambea08} and \citet{SL13}.
For \object{NGC~3184} and \object{NGC~5194} we find evidence for \mbox{\HI} supershells 
close to \HII 
regions in the spiral arms. There is no clear trend where the \HI supershells
are located with respect to the \HII regions in \object{NGC~5194}, whereas more supershells 
appear on the trailing edge for \object{NGC~3184}. 

In the case of \object{NGC~628} (Fig.~\ref{f:NGC628}), \citet{HR15} 
map \HII regions for two arms, 'A' and 'B', and inside corotation, \HI supershells
are indeed found close to and overlapping with the \HII regions, preferentially at their leading edges. 
Especially for 
arm 'B', which is located in an \HI rich part of the galaxy, the \HI supershell locations 
relative to the \HII regions change strikingly near the corotation radius:
Inside, three prominent \HI supershells lie towards the leading edge of the \HII arm,
extending over about a quarter of a turn. Only one small supershell is located
at the trailing edge. From about the corotation radius outwards, the \HI supershells
are spread over the widening \HII arm. None is clearly associated with the leading 
or trailing edges.
It is beyond the scope of this article to explain the differences between these 
galaxies. The fact that the effect we postulate is consistent with the data in 
NGC~628 is, however, encouraging.
}

The \al decay time is comparable to the crossing time through
the \HI supershell, and thus we \rv{expect to} observe it \rv{while it crosses the HI supershells}.
A few Myr later, \rv{\al should isotropise, advect} "vertically" into the halo \citep[e.g.,][]{dAB05}, or mix due to interaction with the cavity walls. Most of the
\al has then decayed, and the contribution to the observed $\gamma$-ray
signal is small.

\section{Conclusions}
We interpret the observed \al kinematics in the Galaxy 
as a consequence of superbubble formation propagating 
with the spiral arms and merging of young superbubbles into older \HI supershells,
\rv{with outflows from currently star-forming
regions into the pre-shaped cavities from preceding star-formation towards the
leading edges of spiral arms}.

The model does not rely on independent offsets between young stars and gaseous 
spiral arms, which 
\rv{might be created by other -- not feedback related -- processes and which}
are a matter of ongoing research \citep[compare, e.g., the review by][]{DB14}.

We conclude that \al mainly decays during the first crossing of superbubbles while in the hot phase.
The bulk of \al is therefore not mixing with cold gas on its decay timescale.
\al has however been found in meteorites indicating its presence  in the gas
that formed the Sun 
\citep[e.g.,][]{GM12}. The corresponding fraction of \al required to mix into a 
star-forming cloud during the decay timescale is, however, small
\citep{Vasilea13}, and  would hardly affect our model.



\begin{acknowledgements}
This research was supported by the cluster of excellence ``Origin
  and Structure of the Universe'' 
and by 
Deutsche Forschungsgemeinschaft
under DFG project number  PR 569/10-1  in the context of the
Priority Program 1573  “Physics of the Interstellar Medium. 
K.\ K.\ was supported by CNES.
\end{acknowledgements}

%
%
%
%
\bibliographystyle{aa}
\bibliography{/Users/mkrause/texinput/references}
\end{document}